\documentclass[conference]{IEEEtran}

\usepackage{cite}
\usepackage{amsmath,amssymb,amsfonts}
\usepackage{algorithm}
\usepackage{algpseudocode}
\usepackage{graphicx}
\usepackage{textcomp}
\usepackage{booktabs}
\usepackage{multirow}
\usepackage{tabularx}
\usepackage{xcolor}
\usepackage[hidelinks]{hyperref}

\newcommand{\rev}[1]{\textcolor{black}{#1}}

\begin{document}

\title{PACT: Post-route Agentic Checkpoint Tuning for
FPGA Timing Closure}

\IEEEoverridecommandlockouts
\author{
\IEEEauthorblockN{
Huan Lin\textsuperscript{*},
Kunlong Li\textsuperscript{*},
Lingli Wang,
and Zhiang Wang
}
\IEEEauthorblockA{
\textit{College of Integrated Circuits and Micro-Nano Electronics} \\
\textit{Fudan University}, Shanghai, China \\
\{huanlin24, klli24\}@m.fudan.edu.cn,
\{llwang, zhiangwang\}@fudan.edu.cn
}
\thanks{\textsuperscript{*}Huan Lin and Kunlong Li contributed equally to this work.}
}

\IEEEaftertitletext{\vspace{-2.0\baselineskip}}

\maketitle

\begin{abstract}
Late-stage FPGA \emph{timing closure} often starts from an implemented
design whose remaining violations are visible in timing reports.
Engineering change order (ECO) optimization is a standard mechanism for applying localized changes to such designs without restarting the full implementation flow.
Automating post-route ECO optimization remains challenging.
A post-route
change must improve timing without violating routing legality, hold or
pulse-width timing constraints, checkpoint replayability
or functional equivalence.
\rev{This paper presents PACT, a
\emph{Post-route Agentic Checkpoint Tuning} framework
for Vivado design checkpoints (DCP).}
\rev{PACT represents post-route tuning as
\emph{validation-gated transitions} between accepted
and candidate checkpoints. From checkpoint-derived evidence, an agent
selects
localized backend actions through a profile-driven recipe planner over
typed Vivado and RapidWright
skills, and probes tool behavior in isolated workspaces.}
PACT records each trial as an evidence-gated case to guide candidate generation and suppress unsafe, unsupported or ineffective actions.
Across 35 UltraScale+ post-route checkpoints, PACT
improves \emph{validation-clean} $F_{\max}$ (maximum operating
frequency) by a geometric mean of
$+22.30\%$ over the original implementations, compared with
$+15.14\%$ for DATuner and $+9.78\%$ for the Codex Agent. On
shared designs, PACT achieves these gains $6.4\times$ faster than the
uncapped DATuner and at an average token cost of only \$0.16 per DCP ($24.5\times$ lower than the free-form Codex Agent). 
The source code is available in an anonymous repository \cite{code}.

\end{abstract}


\section{Introduction}
\label{sec:introduction}

Field-programmable gate arrays (FPGAs) provide flexible acceleration by
mapping application-specific parallelism onto reconfigurable hardware.
Their performance depends not only on the source design but also on
backend implementation. During implementation, logic optimization,
placement, routing and physical optimization together determine the
final signoff frequency. For large or timing-critical designs, engineers
reach timing closure through repeated backend iterations: they inspect
timing reports, adjust constraints or implementation strategies, and
rerun expensive place-and-route steps until the design meets its target
frequency~\cite{amd_ug904_2025,amd_ug906_2025,amd_ug949_2025}. This
process is costly because each full-flow rerun may discard a legal
placed-and-routed state that already contains useful timing, placement
and routing information. Late-stage and incremental methods therefore
aim to preserve this state while repairing the remaining timing
violations.

Engineering change order (ECO) is a standard late-stage mechanism for
modifying an existing implemented design. An ECO applies a small
change after major implementation steps have finished,
rather than rebuilding the design from scratch. Its goal is to fix a
remaining functional, timing or physical issue with limited
perturbation and renewed
validation~\cite{jiang_ir_aware_eco_2024,azadi_graph_based_eco_2026,
rapidwright_eco_tutorial_2025}. In FPGA flows, this model can be
realized through post-route checkpoints, where a Vivado Design
Checkpoint (DCP) preserves the implemented netlist, constraints,
placement, routing and physical context. With Vivado and RapidWright
support, ECO-style post-route checkpoint tuning can inspect, edit,
route and validate an implemented FPGA state instead of blindly
rerunning the full
flow~\cite{amd_ug904_2025,rapidwright_documentation_2025,
zhou_rwroute_2022}.

This ECO-style formulation also exposes a difficult automation
problem. Post-route ECO tuning has traditionally relied on
expert-driven iteration: engineers read timing and physical evidence,
infer the cause of a violation, apply a local fix and validate the
modified checkpoint. The available actions are local, fine-grained and
highly design-dependent. Typical actions involve localized physical
optimization, path-specific repairs, net rerouting, cell replication or
other checkpoint-level changes. The modified checkpoint must then
preserve routing legality, timing consistency and checkpoint usability:
an edit that improves one critical path may degrade another, and a
locally valid change may still produce routing conflicts, constraint
mismatches or non-replayable behavior. \textbf{
The central challenge is to automate the time-consuming and expert-driven ECO process without compromising the rigorous validation required for implemented FPGA states.}

Existing FPGA flow-tuning methods automate a different level of the
problem. Prior work has explored scripted flows, Bayesian optimization,
multi-armed bandits and learning-assisted autotuning for FPGA
implementation recipes~\cite{amd_ug894_2025,amd_ug938_2025,
kapre_intime_2015,xu_datuner_2017,ustun_lamda_2019}. These methods can
reduce manual effort when the search space is defined by directives,
seeds, strategies or tool parameters. However, they usually treat the
backend flow as a black-box mapping from settings to final quality of
results. They do not directly decide which local path, net, cell or
region should be modified inside an already routed checkpoint. They
also do not maintain a validation record for attempted ECO-style
actions. This limits their ability to exploit the detailed evidence
available in an implemented design.

LLM-based EDA agents offer a promising decision layer for this setting.
Recent systems show that large language models can read design reports,
retrieve domain knowledge, invoke EDA tools and refine decisions across
iterations~\cite{pan_survey_llm_eda_2025,liu_chipnemo_2023,
ghose_orfs_agent_2025,ouyang_retrieve_schedule_reflect_2026,
nainani_timing_analysis_agent_2025}. These abilities match the
evidence-rich nature of post-route tuning. However, general tool-using
agents are not sufficient by themselves. Post-route ECO actions must be
tied to a specific accepted checkpoint, executed in isolation,
validated before promotion, and recorded when they fail. Without this
state discipline, an agent may repeat ineffective actions, modify the
wrong design state or accept an unsafe checkpoint.

\rev{This paper presents PACT, a post-route agentic checkpoint-tuning
framework for FPGA.} PACT turns ECO-style checkpoint
editing into a closed-loop optimization process over implemented FPGA
states. At each iteration, an agent reads checkpoint-derived timing and
physical evidence, selects a structured backend action, executes it
using a typed Vivado or RapidWright skill, and promotes the result
only after validation. A profile-driven recipe planner maps timing-profile evidence
to localized optimization actions. A constrained backend
harness isolates exploration, enforces rollback and records attempted
cases within each run. This design combines the flexibility of
LLM-based reasoning with the execution discipline required for
post-route ECO optimization.
Our main contributions are summarized as follows.

\begin{itemize}
\item We introduce an agentic ECO-style post-route optimization
framework for FPGA timing closure. The framework operates directly on
implemented DCPs and converts local backend edits into validation-gated
checkpoint transitions.
Upon acceptance of the paper, the source code will be open-sourced to support reproducibility and facilitate further research and development.

\item We pair a \emph{profile-driven recipe planner} with a typed
library of Vivado and RapidWright skills. The planner maps a
checkpoint's timing profile to effective localized backend actions.

\item We design a \emph{constrained backend harness} for validation-gated execution. The harness isolates exploration, promotes only
validated actions, and suppresses unsafe or repeated trials with a
within-run case record.

\item We evaluate PACT on 35 UltraScale+ post-route checkpoints. PACT
achieves a geometric-mean validation-clean $F_{\max}$ improvement of
$+22.30\%$, compared with $+15.14\%$ for DATuner and $+9.78\%$ for a
free-form Codex Agent.
\end{itemize}

The remainder of this paper is organized as follows.
Section~\ref{sec:related-work} reviews related work and formalizes the
post-route checkpoint tuning problem. \rev{Section~\ref{sec:proposed-method}
presents the PACT framework.} Section~\ref{sec:evaluation} describes the
implementation and evaluation. Section~\ref{sec:conclusion} concludes
the paper.






\section{Related Work}
\label{sec:related-work}

This section reviews four areas most relevant to PACT: FPGA backend
and checkpoint infrastructure, ECO
optimization, LLM-based EDA agents and automated FPGA flow tuning.
It then formulates post-route checkpoint tuning as a
validation-gated state-transition problem
(Section~\ref{sec:problem-formulation}).
Table~\ref{tab:feature_comparison} compares representative work
along the dimensions relevant to PACT.

\begin{table}[t]
\centering
\caption{Feature comparison of representative FPGA
tuning, backend CAD and EDA-agent work.}
\label{tab:feature_comparison}
\scriptsize
\setlength{\tabcolsep}{1.6pt}
\renewcommand{\arraystretch}{1.05}
\newcommand{\cmark}{$\checkmark$}
\newcommand{\xmark}{--}

\resizebox{0.98\columnwidth}{!}{
\begin{tabular}{@{}lcccccccccc@{}}
\toprule
\textbf{Work} &
\rotatebox{60}{\textbf{FPGA}} &
\rotatebox{60}{\textbf{QoR}} &
\rotatebox{60}{\textbf{Search}} &
\rotatebox{60}{\textbf{Tool}} &
\rotatebox{60}{\textbf{Learn.}} &
\rotatebox{60}{\textbf{Agent}} &
\rotatebox{60}{\textbf{Phys.}} &
\rotatebox{60}{\textbf{Impl.}} &
\rotatebox{60}{\textbf{Know.}} &
\rotatebox{60}{\textbf{Safe}} \\
\midrule
InTime~\cite{kapre_intime_2015}
& \cmark & \cmark & \cmark & \cmark & \cmark & \xmark & \xmark &
\xmark & \xmark & \xmark \\
DATuner~\cite{xu_datuner_2017}
& \cmark & \cmark & \cmark & \cmark & \cmark & \xmark & \xmark &
\xmark & \xmark & \xmark \\
LAMDA~\cite{ustun_lamda_2019}
& \cmark & \cmark & \cmark & \cmark & \cmark & \xmark & \xmark &
\xmark & \xmark & \xmark \\
RLPlace~\cite{elgammal_rlplace_2022}
& \cmark & \cmark & \cmark & \xmark & \cmark & \xmark & \cmark &
\xmark & \xmark & \xmark \\
RWRoute~\cite{zhou_rwroute_2022}
& \cmark & \cmark & \xmark & \xmark & \xmark & \xmark & \cmark &
\xmark & \xmark & \xmark \\
AutoChip~\cite{thakur_autochip_2023}
& \xmark & \xmark & \cmark & \cmark & \cmark & \cmark & \xmark &
\xmark & \xmark & \xmark \\
VerilogCoder~\cite{ho_verilogcoder_2025}
& \xmark & \xmark & \cmark & \cmark & \cmark & \cmark & \xmark &
\xmark & \cmark & \xmark \\
TimelyHLS~\cite{mashnoor_timelyhls_2025}
& \cmark & \cmark & \cmark & \cmark & \cmark & \cmark & \xmark &
\xmark & \cmark & \xmark \\
LAAFD~\cite{moraru_laafd_2026}
& \cmark & \cmark & \cmark & \cmark & \cmark & \cmark & \xmark &
\xmark & \xmark & \xmark \\
ORFS-agent~\cite{ghose_orfs_agent_2025}
& \xmark & \cmark & \cmark & \cmark & \cmark & \cmark & \cmark &
\xmark & \xmark & \xmark \\
RSR~\cite{ouyang_retrieve_schedule_reflect_2026}
& \xmark & \cmark & \cmark & \cmark & \cmark & \cmark & \cmark &
\xmark & \cmark & \xmark \\
TA Agent~\cite{nainani_timing_analysis_agent_2025}
& \xmark & \cmark & \xmark & \cmark & \cmark & \cmark & \xmark &
\xmark & \cmark & \xmark \\
\midrule
\textbf{PACT}
& \cmark & \cmark & \cmark & \cmark & \cmark & \cmark & \cmark &
\cmark & \cmark & \cmark \\
\bottomrule
\end{tabular}
}

\vspace{0.5mm}
\begin{minipage}{0.98\columnwidth}
\scriptsize
\emph{Notes.}
\textbf{QoR}: timing/performance/QoR-oriented
optimization.
\textbf{Search}: automated or iterative exploration.
\textbf{Tool}: uses EDA-tool feedback, reports,
simulation or implementation results as optimization
signals.
\textbf{Learn.}: machine learning (ML), reinforcement learning (RL)
and LLM-based decision making.
\textbf{Agent}: agentic LLM workflow.
\textbf{Phys.}: backend or physical-design-level
action.
\textbf{Impl.}: optimization on an already implemented
post-route checkpoint or state.
\textbf{Know.}: explicit knowledge, retrieval, reflection
or memory.
\textbf{Safe}: validation-gated optimization with rollback.
\end{minipage}
\end{table}

\subsection{FPGA Backend and Checkpoint Infrastructure}
\label{sec:backend-cad-checkpoint-substrates}
\label{sec:rapidwright-based-frameworks}

Commercial FPGA implementation flows provide the core backend
operations needed for timing closure, including optimization,
placement, physical optimization and routing, together with signoff
analysis and checkpoint-based implementation
support~\cite{amd_ug904_2025,amd_ug906_2025,amd_ug949_2025,
amd_ug894_2025,amd_ug938_2025}. RapidWright complements these
capabilities by exposing APIs for
reading, writing and manipulating Vivado DCPs~\cite{rapidwright_documentation_2025}. It also supports
inspection of implemented designs and timing-driven
routing~\cite{zhou_rwroute_2022}. Together, these capabilities enable
post-route checkpoint editing without rebuilding the design from RTL.
PACT uses this infrastructure to execute backend edits, but focuses on
how an optimizer manages the checkpoint-tuning process.

\subsection{Engineering Change Order Optimization}
\label{sec:eco-related-work}

ECO methods address late-stage functional,
timing or physical problems through localized modifications to an
implemented design. For FPGA checkpoints, such
modifications can include
localized physical optimization, net rerouting, inserted logic, cell
replication and checkpoint repair. Despite their different
mechanisms, these methods share the goal of improving signoff quality
while minimizing perturbation to the existing implementation.

Recent work has introduced learning-based decision making into ECO
optimization. Jiang et al.~\cite{jiang_ir_aware_eco_2024} combine
reinforcement
learning with gate sizing for IR-aware timing repair.
Azadi et al.~\cite{azadi_graph_based_eco_2026} use graph edits to
generate compact ECO patches at the high-level synthesis (HLS)
intermediate-representation level. These studies demonstrate the value
of targeted, state-preserving changes, but focus on specific
abstraction levels or repair mechanisms.

For FPGA designs, checkpoint-based frameworks provide mechanisms for
applying ECO-style changes to implemented designs. RapidWright, for
example, supports inserting and routing new logic in an existing
placed-and-routed design without full
reimplementation~\cite{rapidwright_eco_tutorial_2025}. However, such frameworks mainly provide APIs for
applying edits to an implemented checkpoint.
They do not provide an automated strategy for selecting
ECO actions or comparing candidate checkpoints.
Post-route ECO tuning therefore still relies heavily on
expert-driven iteration. Engineers inspect timing
evidence, choose a local edit, validate the candidate checkpoint and
decide whether to keep or discard it.

\subsection{LLM-Based EDA Agents}
\label{sec:llm-based-eda-agents}

Recent LLM-based EDA agents~\cite{pan_survey_llm_eda_2025,
liu_chipnemo_2023,thakur_autochip_2023,ho_verilogcoder_2025,
mashnoor_timelyhls_2025,moraru_laafd_2026} demonstrate that language
models can interpret design reports, retrieve domain knowledge, invoke
EDA tools and refine decisions through iterative feedback. Most of
these systems focus on RTL generation, functional debugging or HLS
optimization, and therefore operate before physical implementation.
Backend-oriented agents~\cite{ghose_orfs_agent_2025,
ouyang_retrieve_schedule_reflect_2026,
nainani_timing_analysis_agent_2025} extend LLM-based reasoning to
physical-design reports, timing analysis and tool-flow orchestration.
However, they mainly target ASIC-oriented flows or analysis tasks.
\rev{None of the reviewed systems directly optimizes post-route FPGA checkpoints.}

\subsection{Automated FPGA Flow Tuning}
\label{sec:automated-fpga-flow-tuning}

Automated FPGA design-space exploration reduces manual timing-closure
effort by searching over implementation strategies, directives, seeds
and tool parameters. InTime, DATuner and
LAMDA~\cite{kapre_intime_2015,xu_datuner_2017,ustun_lamda_2019} use
learning or search to prioritize expensive implementation trials and
improve timing convergence. Other work formulates placement or routing
as sequential
decision-making problems~\cite{elgammal_rlplace_2022,
esmaeili_guiding_fpga_detailed_placement_2022,
farooq_efficient_fpga_routing_2021}. These approaches either tune
global implementation settings or optimize decisions within a fixed CAD
engine. They do not select localized modifications to paths,
nets, cells or regions in an already routed FPGA checkpoint 
based on semantic information extracted from EDA reports.

In summary, prior work provides checkpoint-manipulation mechanisms,
ECO-style local repair, automated flow search and LLM-based EDA
reasoning. However, these ingredients have not been
combined at the post-route FPGA checkpoint level. A
suitable optimizer must select local ECO actions from timing evidence,
execute them through constrained backend tools, validate the resulting
checkpoint and remember failed or unsafe attempts. PACT
addresses this gap with an agentic, validation-gated checkpoint-tuning
formulation, which we formalize next.

\subsection{Problem Formulation}
\label{sec:problem-formulation}

Given an initial post-route FPGA checkpoint $D_0$, post-route backend
optimization applies a sequence of local actions to implemented design
states. Applying an action $a_t$ to the current checkpoint $D_t$
produces
\begin{equation}
    D_{t+1} = a_t(D_t).
\label{eq:pact-transition}
\end{equation}
Unlike flow-level tuning, these actions modify an already
placed-and-routed design rather than restarting implementation from
RTL or synthesis.

Within a tuning budget $B$, the objective is to find a reachable
checkpoint with the highest operating frequency:
\begin{equation}
\begin{aligned}
D^\star =& \arg\max_{D \in \mathcal{R}_B(D_0)}
    \quad  F_{\max}(D) \\
\mathrm{s.t.}\quad
    & \mathrm{Validate}(D,C)=\mathrm{true},
\end{aligned}
\label{eq:pact-objective}
\end{equation}
where $\mathcal{R}_B(D_0)$ is the set of checkpoints reachable within
the budget. The validation set $C$ covers routing legality, hold and
pulse-width timing and checkpoint usability; functional equivalence is
ensured by construction (Section~\ref{sec:action-library}), not checked
in $C$. The main challenge is to select local actions that improve
timing without invalidating the implemented design. PACT addresses
this problem as described in the next section.

\section{Proposed Method}
\label{sec:proposed-method}

\rev{This section presents PACT, an agentic framework for post-route
FPGA ECO.} It first gives a system overview
(Section~\ref{sec:system-overview}), then describes ECO diagnosis and
profile-driven planning (Section~\ref{sec:run-state}), the structured
ECO action space (Section~\ref{sec:action-library}) and the constrained
execution loop (Section~\ref{sec:optimization-loop}), and finally the
validation, acceptance and termination rules
(Section~\ref{sec:validation-gates}).

\subsection{System Overview}
\label{sec:system-overview}

\begin{figure*}[t]
\centering
\includegraphics[width=0.9\linewidth]{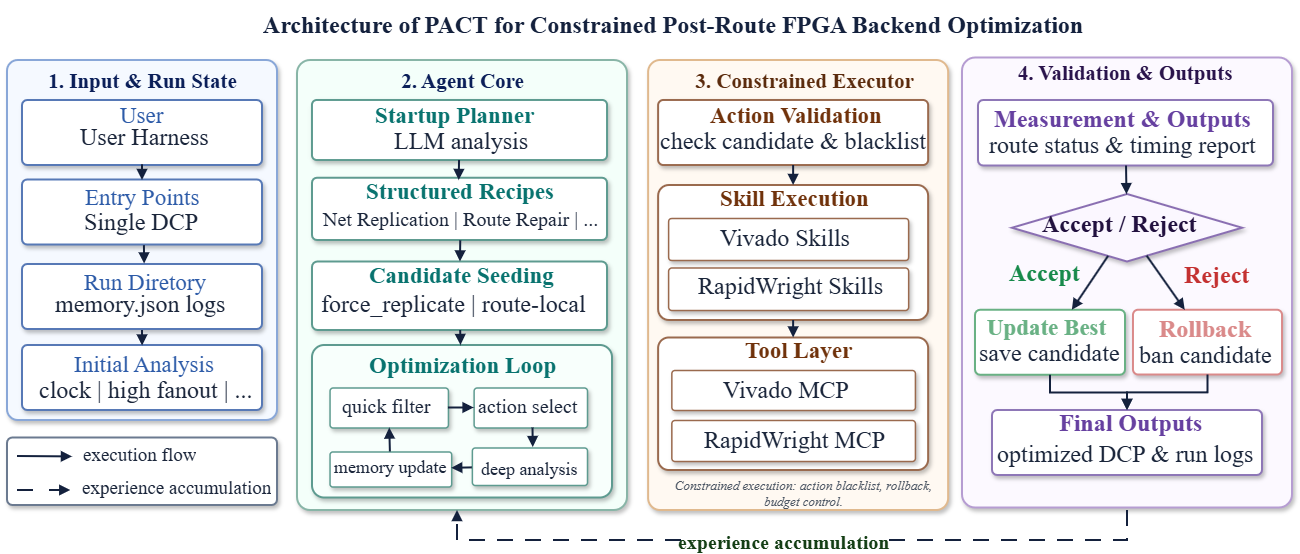}
\caption{System overview of PACT: a checkpoint-derived run state feeds
an agent core that proposes structured candidate actions, a
constrained executor runs only admissible actions through Vivado or
RapidWright skills and the validation stage accepts
only
validation-clean improvements.}
\label{fig:pact-overview}
\end{figure*}

As Figure~\ref{fig:pact-overview} shows, a single iteration works
through four stages over the implemented design. The diagnosis stage extracts timing,
routing and physical evidence from the accepted checkpoint and
summarizes it as a timing profile, a structured description of the
current timing failure. The agentic planner uses this profile to rank
candidate actions from a typed ECO action library
(Table~\ref{tab:actions}). PACT then selects the top-ranked action and
evaluates it through the matching Vivado or RapidWright skill. The
validation stage promotes the result only when it is a fully routed
checkpoint that improves setup timing; otherwise PACT keeps the accepted
checkpoint. Throughout, PACT separates reasoning from execution. The
agent selects and parameterizes actions but never edits the checkpoint
itself; every backend change passes through the constrained executor
under the same validation policy.

\subsection{ECO Diagnosis and Profile-Driven Planning}
\label{sec:run-state}

A tuning run receives an input checkpoint $D_0$, a validation
policy $C$ and a budget $B$. PACT regenerates its reports from the
current checkpoint so that the evidence always matches the accepted
state. Global setup metrics come from \texttt{report\_timing\_summary},
critical-path detail from \texttt{report\_timing}, routing status from
\texttt{report\_route\_status} and fanout from Vivado net queries.
RapidWright supplies object-level detail when finer inspection is
needed. PACT condenses this evidence into the timing profile, which
records worst and total negative slack, critical start and end points,
logic and routing delay, high-fanout nets and routing problems. The run
state also
holds the accepted checkpoint, its report bundle, the remaining budget,
the candidate queue and a within-run evidence record.

The profile-driven planner links timing bottlenecks to executable
backend actions. The recipe library $P$ maps common bottlenecks to the
action families in Table~\ref{tab:actions}. For example, high fanout
together with route-dominated delay suggests driver replication,
whereas excessive routing delay on a critical net suggests local
rerouting. These mappings provide candidate actions rather than a
fixed execution schedule.

The agent ranks and parameterizes the applicable candidates from
the complete run state. When the diagnosis is incomplete, it may issue
non-mutating diagnostics, read-only probes that gather more evidence
without changing the accepted checkpoint. The within-run record $M$
suppresses exact repeats and deprioritizes actions that already failed
on the same accepted state. Because planning and execution communicate
only through this structured interface, the agentic planner is
replaceable: the same action library, budget and validation policy run
unchanged under either the agent or a deterministic rule-based
selector.

\subsection{Structured ECO Action Space}
\label{sec:action-library}

PACT's action space comprises the nine families in
Table~\ref{tab:actions}, partitioned into local ECO actions, global
fallbacks and non-mutating diagnostics. Local actions repair a specific
path, net, cell group or region. Global fallbacks broaden the search
when local evidence is weak or local candidates are exhausted.
Diagnostics gather evidence but do not themselves improve timing.

\begin{table}[t]
\centering
\scriptsize
\setlength{\tabcolsep}{3pt}
\caption{The PACT action families and the evidence that triggers them.
Backends: V~=~Vivado and RW~=~RapidWright.}
\label{tab:actions}
\begin{tabularx}{\columnwidth}{@{}>{\raggedright\arraybackslash}p{2cm}
>{\raggedright\arraybackslash}X
>{\centering\arraybackslash}p{0.7cm}@{}}
\toprule
\textbf{Family} & \textbf{Trigger and operation} & \textbf{Tool} \\
\midrule
\multicolumn{3}{@{}l}{\emph{Local ECO actions}} \\
Fanout replication & High fanout, route-dominated delay; replicate driver, split sinks, repair routing. & V, RW \\
Local rerouting & High route delay on critical nets; unroute and reroute a bounded window. & V \\
Local re-placement & Cell-local path delay; move cells or endpoints, repair affected routes. & V, RW \\
Logic restructuring & LUT-dominated critical cone; LUT-cone or input-pin transform. & V, RW \\
Floorplanning & Congestion or long inter-region paths; adjust a pblock, reimplement region. & V, RW \\
\midrule
\multicolumn{3}{@{}l}{\emph{Global fallback actions}} \\
Directive sweep & Weak or diffuse evidence; sweep post-route opt and routing directives. & V \\
Re-place/route & Local candidates exhausted; timing-driven place-and-route from current DCP. & V \\
Clock retargeting & Need stronger pressure; tighten temporarily, validate under original clock. & V \\
\midrule
\multicolumn{3}{@{}l}{\emph{Non-mutating diagnostics}} \\
Control \& probing & Import, report or inspect objects; never mutates the checkpoint. & V, RW \\
\bottomrule
\end{tabularx}
\end{table}

A concrete action is a typed instance of one family, fixing its
operation type, target objects, parameters, backend, preconditions,
required checks and estimated cost. A replication action, for example,
names the driver net and sink partition, and a rerouting action the
selected net window. The executor can therefore inspect an action
before execution and tie the result to it afterward. The library
excludes intentional functional changes by construction: every mutating
action is a logic- or timing-equivalent transform, and PACT never counts
relaxed timing constraints as an improvement. All reported timing
therefore uses the original clock and timing-exception constraints.

\subsection{Constrained Execution and Optimization Loop}
\label{sec:optimization-loop}

Before execution, PACT screens the selected action for
admissibility. 
\rev{An action is rejected if its type is unsupported, required targets
are missing, preconditions fail, its estimated cost exceeds the remaining
budget, or $M$ marks it as duplicate or contraindicated.}
A proposal that fails these checks never reaches the backend
tools. Each supported operation is a typed skill that wraps a controlled
subset of Vivado Tcl or RapidWright calls. The skill runs on a copy of
the accepted checkpoint in an isolated trial directory and returns a
fixed artifact set: an output DCP, timing reports, a route-status
report, tool logs and runtime metadata. The agent therefore sees only
these structured outcomes, with no shell access and no direct write
access to the checkpoint.

Algorithm~\ref{alg:pact} summarizes the optimization loop. The
recipe library $P$ contains the fixed bottleneck-to-action mappings. The
evidence record $M$ contains only information collected during the
current run. The candidate queue contains mutating ECO and fallback
actions. Non-mutating diagnostics update the run state directly and do
not enter the acceptance test. \rev{Algorithm~\ref{alg:pact} proceeds in
three stages. It first validates the input DCP and returns early if the
starting state is invalid
[Lines~\ref{alg:pact:init-memory}--\ref{alg:pact:init-check-end}]. It
then commits $D_0$ as the accepted checkpoint, builds the run state and
seeds the candidate queue
[Lines~\ref{alg:pact:set-best}--\ref{alg:pact:plan-init}]. Each iteration
pops the best action, applies the admissibility filter, and runs an
admissible action in isolation from the accepted checkpoint
[Lines~\ref{alg:pact:loop-start}--\ref{alg:pact:execute}]. A
validation-clean score gain promotes the candidate; otherwise PACT
discards it and logs a rejected case
[Lines~\ref{alg:pact:unavailable-if}--\ref{alg:pact:reject-record}]. After
every trial it replans from the accepted checkpoint, so failed edits
never accumulate, and it returns the best validated checkpoint once the
budget or queue is exhausted
[Lines~\ref{alg:pact:rebuild-state}--\ref{alg:pact:return}].}

\begin{algorithm}[t]
\footnotesize
\caption{Post-route ECO optimization in PACT.}
\label{alg:pact}
\begin{algorithmic}[1]
\Require input DCP $D_0$, validation policy $C$, budget $B$,
recipe library $P$

\State $M \gets \emptyset$ \label{alg:pact:init-memory}
\State $(R_{\mathrm{best}},v_0) \gets \textsc{Evaluate}(D_0,C)$ \label{alg:pact:init-eval}

\If{not $v_0$} \label{alg:pact:init-check}
    \State \Return input validation failure \label{alg:pact:init-return}
\EndIf \label{alg:pact:init-check-end}

\State $D_{\mathrm{best}} \gets D_0$ \label{alg:pact:set-best}
\State $S \gets \textsc{BuildState}
(D_{\mathrm{best}},R_{\mathrm{best}},M,B)$ \label{alg:pact:build-state-init}
\State $Q \gets \textsc{Plan}(S,P,M)$ \label{alg:pact:plan-init}

\While{$\textsc{WithinBudget}(B)$ and $Q \neq \emptyset$} \label{alg:pact:loop-start}
    \State $a \gets \textsc{PopBest}(Q)$ \label{alg:pact:pop-best}

    \If{not $\textsc{Admissible}(a,S,C,M)$} \label{alg:pact:admissible-if}
        \State $M \gets \textsc{RecordSkip}(M,a)$ \label{alg:pact:record-skip}
        \State \textbf{continue} \label{alg:pact:skip-continue}
    \EndIf \label{alg:pact:admissible-end}

    \State $D' \gets
    \textsc{ExecuteIsolated}(D_{\mathrm{best}},a)$ \label{alg:pact:execute}

    \If{$D'$ is unavailable} \label{alg:pact:unavailable-if}
        \State $M \gets \textsc{RecordFailure}(M,a)$ \label{alg:pact:record-failure}
    \Else
        \State $(R',v') \gets \textsc{Evaluate}(D',C)$ \label{alg:pact:evaluate-candidate}

        \If{$v'$ and
        $\textsc{Score}(R') >
        \textsc{Score}(R_{\mathrm{best}})+\epsilon$} \label{alg:pact:accept-if}
            \State $D_{\mathrm{best}} \gets D'$ \label{alg:pact:update-best}
            \State $R_{\mathrm{best}} \gets R'$ \label{alg:pact:update-report}
            \State $M \gets
            \textsc{RecordAccept}(M,a,R')$ \label{alg:pact:record-accept}
        \Else \label{alg:pact:reject-else}
            \State $\textsc{Discard}(D')$ \label{alg:pact:discard}
            \State $M \gets
            \textsc{RecordReject}(M,a,R')$ \label{alg:pact:reject-record}
        \EndIf
    \EndIf

    \State $S \gets \textsc{BuildState}
    (D_{\mathrm{best}},R_{\mathrm{best}},M,B)$ \label{alg:pact:rebuild-state}
    \State $Q \gets \textsc{UpdatePlan}(Q,S,P,M)$ \label{alg:pact:update-plan}
\EndWhile \label{alg:pact:loop-end}

\State \Return $D_{\mathrm{best}}$, $R_{\mathrm{best}}$ and $M$ \label{alg:pact:return}
\end{algorithmic}
\end{algorithm}

\subsection{Validation, Acceptance and Termination}
\label{sec:validation-gates}

Before scoring, PACT checks that a candidate checkpoint is valid. A
valid output DCP is reopenable, fully routed and free of routing errors,
and passes the required hold, pulse-width, minimum-period and
design-consistency checks.

Among valid candidates, PACT accepts a trial $D'$ only when its
setup worst slack improves by at least $\epsilon=0.5$~ps,
$s(D') > s(D_{\mathrm{best}}) + \epsilon$, where $s(\cdot)$ is the worst
negative slack (WNS) under the design's original timing constraints, so a temporarily
tightened clock never counts as an improvement. This rule keeps the
committed trajectory monotonic, while exploration stays diverse: after
each trial, PACT updates the timing profile and reranks both local ECO
actions and global fallbacks. PACT therefore targets a strong validated
result within the budget, not a globally optimal one.

Rejected candidates are discarded, while their actions and
validation results are recorded to avoid repeated failures. The run
terminates when its time, token or trial budget is exhausted, or when
no admissible candidate remains. PACT returns the best validated
checkpoint and its reports.

\section{Evaluation}
\label{sec:evaluation}

This section first describes how PACT is implemented around Vivado
and RapidWright (Section~\ref{sec:implementation}) and the benchmark
suite (Section~\ref{sec:benchmarks}), and then reports the experimental
setup, baselines, main results, and an ablation and generalization
analysis.

\subsection{Implementation}
\label{sec:implementation}

We implement PACT as an agentic post-route optimization framework
built around Vivado and RapidWright. Each supported operation of
Section~\ref{sec:proposed-method} is realized as a typed skill that
renders the action parameters into the underlying Vivado Tcl or
RapidWright call, executed in isolation as described in
Section~\ref{sec:optimization-loop}. The two backends play
complementary roles: Vivado performs checkpoint loading, timing and
route report extraction, post-route optimization, checkpoint writing
and signoff validation, while RapidWright provides implemented-design
inspection and object-level analysis when lower-level checkpoint access
is needed.

\subsection{Benchmarks}
\label{sec:benchmarks}

\begin{table}[t]
\centering
\scriptsize
\setlength{\tabcolsep}{3pt}
\caption{Benchmark characteristics, grouped into the four categories.
Test designs are marked with~$\dagger$. $F_{\max}$ is reported in MHz and extracted from the
original DCPs.}
\label{tab:benchmark-characteristics}
\begin{tabular}{>{\raggedright\arraybackslash}p{2.8cm}rrrrr}
\toprule
Benchmark & LUTs & FFs & DSPs & BRAMs & $F_{\max}$ \\
\midrule
\multicolumn{6}{@{}l}{\textit{(1) Classic, compact and HLS-style benchmarks}}\\
AMD Mini-ISP~\cite{xilinx_fpl26_optimization_contest_2026} & 3k & 4k & 40 & 12 & 307.1 \\
LogicNets JSCL~\cite{xilinx_fpl26_optimization_contest_2026} & 31k & 2k & 0 & 0 & 403.6 \\
Rosetta 3D Rendering~\cite{xilinx_fpl26_optimization_contest_2026} & 14k & 5k & 3 & 0 & 270.9 \\
Rosetta Digit Recognition~\cite{xilinx_fpl26_optimization_contest_2026} & 23k & 23k & 0 & 161 & 367.0 \\
Rosetta Optical Flow~\cite{xilinx_fpl26_optimization_contest_2026} & 34k & 37k & 42 & 61 & 324.9 \\
Rosetta Spam Filter~\cite{xilinx_fpl26_optimization_contest_2026} & 5k & 13k & 224 & 3 & 437.4 \\
VTR MCML~\cite{xilinx_fpl26_optimization_contest_2026} & 43k & 15k & 105 & 142 & 62.2 \\
AES~\cite{secworks_aes} & 3.4k & 3.0k & 0 & 0 & 334.0 \\
TCAM~\cite{vtr_benchmarks} & 156 & 281 & 0 & 0 & 561.8 \\
VexRiscv Re-place~\cite{vexriscv_repo} & 2k & 1k & 4 & 6 & 310.2 \\
OpenRISC 1200$^\dagger$~\cite{openrisc_or1200} & 4.6k & 2.5k & 4 & 4 & 209.0 \\
\midrule
\multicolumn{6}{@{}l}{\textit{(2) Compute-intensive datapath designs}}\\
FINN RadioML~\cite{xilinx_fpl26_optimization_contest_2026} & 74k & 46k & 0 & 25 & 284.9 \\
FSA Systolic Array~\cite{vca_epfl_fsa} & 373.8k & 29.2k & 256 & 0 & 82.8 \\
Double-Clock FFT-L~\cite{zipcpu_dblclockfft} & 10.8k & 26.2k & 0 & 32 & 442.5 \\
Double-Clock FFT-M~\cite{zipcpu_dblclockfft} & 5.4k & 13.1k & 0 & 16 & 456.4 \\
Double-Clock FFT-S~\cite{zipcpu_dblclockfft} & 1.4k & 3.3k & 0 & 4 & 477.1 \\
RaygenTop~\cite{vtr_benchmarks} & 986 & 1.7k & 9 & 0 & 451.3 \\
Verilog CORDIC-L~\cite{opencores_cordic} & 23.4k & 23.9k & 0 & 0 & 456.6 \\
Verilog CORDIC-M~\cite{opencores_cordic} & 11.7k & 12.0k & 0 & 0 & 473.7 \\
Verilog CORDIC-S~\cite{opencores_cordic} & 750 & 823 & 0 & 0 & 682.1 \\
NVDLA Large$^\dagger$~\cite{nvidia_nvdla_hw} & 241.5k & 30.6k & 0 & 0 & 158.1 \\
NVDLA XL~\cite{nvidia_nvdla_hw} & 272.6k & 33.7k & 0 & 0 & 152.5 \\
NVDLA Basic~\cite{nvidia_nvdla_hw} & 30.2k & 722 & 0 & 0 & 189.9 \\
\midrule
\multicolumn{6}{@{}l}{\textit{(3) Large system-level designs}}\\
BOOM SoC~\cite{xilinx_fpl26_optimization_contest_2026} & 227k & 98k & 61 & 161 & 48.2 \\
CoreScore-500~\cite{xilinx_fpl26_optimization_contest_2026} & 100k & 120k & 0 & 250 & 344.2 \\
ISPD16 Example2~\cite{xilinx_fpl26_optimization_contest_2026} & 289k & 234k & 200 & 384 & 107.6 \\
AXIS Fabric$^\dagger$~\cite{corundum_repo} & 76.5k & 68.2k & 0 & 144 & 179.1 \\
AXIS Fabric-Hard$^\dagger$~\cite{corundum_repo} & 229.8k & 204.6k & 0 & 432 & 140.0 \\
Vortex Xbar Large$^\dagger$~\cite{vortex_repo} & 215.5k & 156.9k & 0 & 0 & 146.8 \\
Vortex Xbar Basic~\cite{vortex_repo} & 41.4k & 27.3k & 0 & 0 & 210.5 \\
\midrule
\multicolumn{6}{@{}l}{\textit{(4) Generated backend-focused checkpoints}}\\
Congestion XL$^\dagger$ & 84.6k & 104.1k & 48 & 64 & 305.2 \\
FFT-CORDIC-L$^\dagger$ & 60.7k & 77.4k & 32 & 48 & 338.4 \\
FPL26 Self-built & 10.7k & 7.9k & 144 & 24 & 235.0 \\
FFT-CORDIC-M & 30.9k & 39.1k & 16 & 24 & 343.9 \\
SoC-L$^\dagger$ & 43.7k & 52.8k & 32 & 32 & 326.9 \\
\bottomrule
\end{tabular}
\end{table}

We collected 35 post-route Vivado design checkpoints targeting the AMD
UltraScale+ \texttt{xcvu3p-ffvc1517-2-e} device. They come from four
sources: public contest
checkpoints~\cite{xilinx_fpl26_optimization_contest_2026}; open-source
RTL projects~\cite{corundum_repo, vca_epfl_fsa, nvidia_nvdla_hw,
vortex_repo, secworks_aes, zipcpu_dblclockfft, openrisc_or1200,
opencores_cordic}; classic academic and HLS-style benchmark
designs~\cite{vtr_benchmarks}; and internally generated stress-test
checkpoints with deliberately tightened clock targets and near-maximal
resource utilization. We designate the
$\dagger$-marked checkpoints in Table~\ref{tab:benchmark-characteristics}
as a held-out test set that PACT optimizes cold, and use the remaining
designs for development. PACT's within-run case record does not carry
across designs. The split therefore isolates generalization of
the action set and recipe planner rather than memorized per-design
recipes.

The benchmark suite spans four design groups:
\begin{itemize}
\item classic, compact and HLS-style benchmarks;
\item compute-intensive datapaths, including FFT, CORDIC,
systolic-array and rendering designs;
\item large system-level designs such as SoCs, processors,
accelerators and network fabrics, with global placement effects and
high-fanout routing;
\item generated backend-focused checkpoints that stress congestion
and mixed-resource paths.
\end{itemize}
Together, these groups test generalization across multiple
post-route timing-failure modes rather than a single narrow pattern.

\subsection{Experimental Setup}
\label{sec:experimental-setup}

We evaluate all methods on post-route Vivado checkpoints targeting the
\texttt{xcvu3p-ffvc1517-2-e} device, using Vivado~2025.1 and
RapidWright~2025.1 as the backend toolchain and GPT-5.5 as the
underlying model for the agentic methods. For fair comparison, all runs share the same resource limit---eight cores and 32~GB RAM on an Ubuntu~22.04 machine with an Intel Core i7-12700 processor---and the same validation policy $C$. The primary metric is the best validation-clean $F_{\max}$ obtained within the tuning budget, where a result is \emph{validation-clean} only when it passes the full policy of Section~\ref{sec:validation-gates}: a reopenable, fully routed DCP with zero routing errors; non-negative hold, pulse-width and min-period slack and no failing endpoints; a setup/$F_{\max}$ measured against the original target clock, with $F_{\max}=1000/(T-\mathrm{WNS})$~MHz for target-clock period $T$; and preserved primary I/O and clock definitions (functional equivalence holds by construction).

The two agentic methods, PACT and the Codex Agent, run under a one-hour
wall-clock budget per design.
\rev{A PACT run rarely uses the full budget. Its loop exits once no
admissible candidate remains: queued actions are exhausted, infeasible
within the budget, or suppressed by the case record. PACT then returns
the best validated checkpoint, so most runs finish well before the cap
(Section~\ref{sec:time-efficiency}).}
DATuner is not subject to this cap and instead terminates on its own
search budget (Section~\ref{sec:baselines}).

Across all experiments, PACT draws only from the bounded
backend-action set of Table~\ref{tab:actions} and commits only
route-legal, validation-clean improvements. All reported $F_{\max}$
values use the original clock and timing-exception constraints, without
constraint relaxation.

\subsection{Baselines}
\label{sec:baselines}

\textbf{DATuner.}
We use DATuner as a search-based autotuning
baseline, representing the class of FPGA tool-flow tuners that
explore a parameterized implementation space using search. We run its
bandit-based space-partitioning algorithm over a $47{,}250$-point
implementation space: Vivado \texttt{opt}, \texttt{place},
\texttt{phys\_opt} and \texttt{route} directives, a fanout limit and a
clock-tightening target factor. We ran place-and-route for each
trial on the same benchmark designs and evaluated it under the same
validation gates as PACT, but without the one-hour wall-clock budget
imposed on the agentic methods.
This favorable setting tests whether PACT can reach competitive
or better validation-clean results more time-efficiently through
structured post-route checkpoint actions.

\begin{figure*}[t]
\centering
\includegraphics[width=0.9\linewidth]{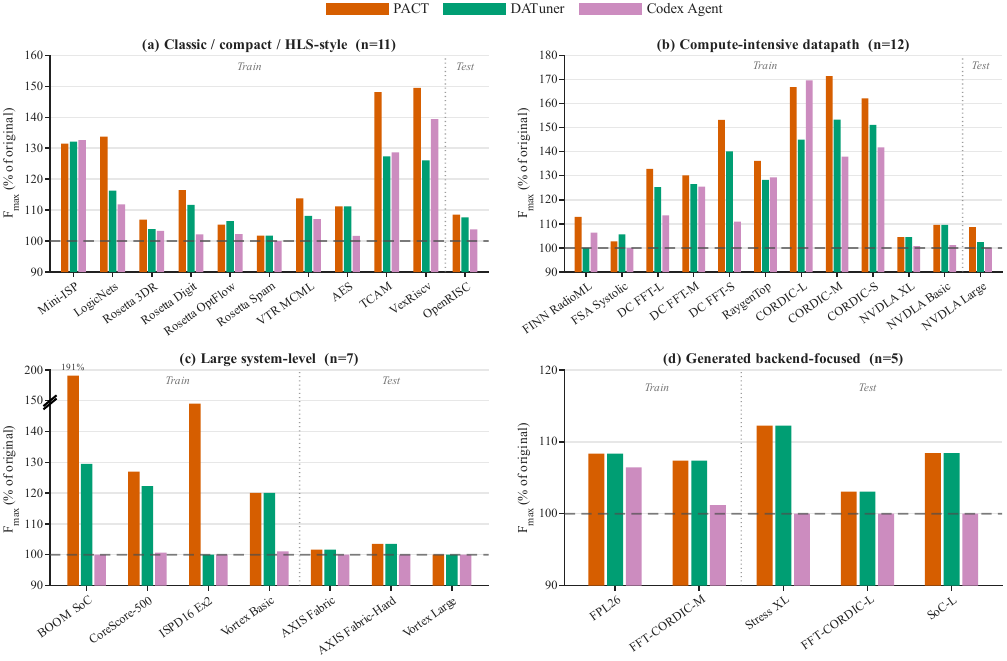}
\caption{Per-design $F_{\max}$ achieved by PACT, DATuner
and the
Codex Agent, normalized to the original post-route checkpoint
(Original${} = 100\%$, dashed). The dotted divider
separates the development
designs and held-out test designs.}
\label{fig:fmax}
\end{figure*}

\textbf{Codex Agent.}
We use Codex command-line interface (CLI)~0.136.0 as a
knowledge-and-tool-matched LLM-agent baseline. \rev{The baseline receives
the same optimization knowledge, Vivado and RapidWright Model Context
Protocol (MCP) servers, reports, and checkpoint inputs as PACT, and runs
GPT-5.5 under the same one-hour budget.}
However, it does not use PACT's constrained execution
harness, structured action representation, validation-gated
acceptance or rollback discipline. This comparison
isolates the
contribution of PACT's harness from the contribution of LLM reasoning,
domain knowledge and tool availability.

\subsection{Main Results}
\label{sec:main-results}

\subsubsection{Frequency Improvement}
\label{sec:frequency-improvement}

Figure~\ref{fig:fmax} compares the best validation-clean $F_{\max}$ obtained by each method, normalized to the original post-route checkpoint. PACT achieves the strongest overall result, improving $F_{\max}$ by a geometric mean of $\mathbf{+22.30\%}$ across the 35-design benchmark suite. In comparison, DATuner improves by $\mathbf{+15.14\%}$ and the Codex Agent by $\mathbf{+9.78\%}$. Per design, PACT obtains the best or tied-best result on 27 of 35 benchmarks, showing that its aggregate gain is not driven by a small number of outliers.

PACT also reaches these results with limited backend exploration. On
average, PACT evaluates only $2.6$ checkpoint-modifying backend actions
per DCP, excluding read-only diagnostic probes. In comparison, DATuner
explores about $12$ full place-and-route trials per design, while the
free-form Codex Agent issues roughly $60$ tool calls per design. Although
these counts reflect different execution granularities, they show that
PACT concentrates the search into a small number of targeted,
validation-gated checkpoint edits.

The largest improvements appear on checkpoints with recoverable timing margin. For example, PACT improves BOOM SoC from $48.2$ to $91.9$~MHz, raises the Verilog CORDIC family by up to $1.71\times$ and improves TCAM from $561.8$ to $831.9$~MHz. On congestion-bound designs, PACT often preserves the original implementation rather than accepting unsafe or marginal edits. Vortex Xbar Large shows this behavior, as all methods find limited additional margin.

\subsubsection{Time Efficiency}
\label{sec:time-efficiency}

\begin{figure}[t]
\centering
\includegraphics[width=0.9\linewidth]{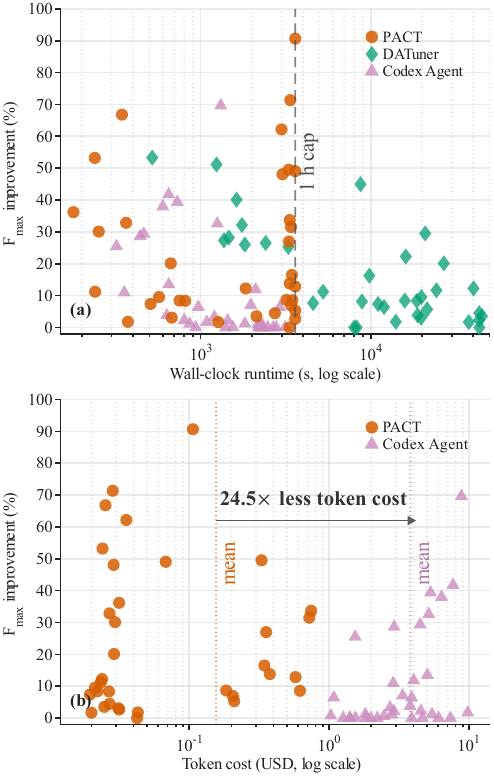}
\caption{(a)~Quality--runtime and (b)~quality--cost tradeoffs:
per-design $F_{\max}$ improvement over the original (\%) versus
wall-clock runtime and token cost for PACT, DATuner and the Codex Agent. Dashed line:
one-hour cap; dotted lines: each method's average per-design cost.}
\label{fig:time}
\end{figure}

PACT achieves its frequency gains within the one-hour wall-clock budget on all 35 designs, with a mean runtime of 36 minutes (${2140}$\,s). In contrast, DATuner is run without this cap, giving the search-based autotuner a favorable setting, yet still requires much longer runtimes: 4.5 hours on average (${16070}$\,s) and up to 12.5 hours (${45162}$\,s). As shown in Figure~\ref{fig:time}(a), DATuner lies substantially to the right of PACT on the runtime axis, running slower on 30 of the 35 shared designs by a geometric mean of $6.4\times$.

The additional runtime does not translate into better quality: on the 35 designs completed by both methods, PACT achieves higher $F_{\max}$ on 26 designs and DATuner on seven, so PACT delivers comparable-or-better timing quality with bounded runtime and substantially lower wall-clock cost. The Codex Agent also fits within the one-hour budget with a lower mean runtime of 25 minutes (${1524}$\,s) but, as noted above, achieves lower frequency gains, leaving PACT with the best time--quality tradeoff among the evaluated methods.

\subsubsection{Token Efficiency}
\label{sec:token-efficiency}

The Codex Agent's free-form harness costs \textbf{\$3.81 per DCP on average}, whereas PACT costs only \textbf{\$0.16 per DCP on average} on the same 35 designs. On a per-DCP basis this is a $24.5\times$ cost reduction for PACT, under the same GPT-5.5 model, one-hour wall-clock budget, optimization knowledge, checkpoint inputs and Vivado/RapidWright tool access---the main difference being the execution harness.

As shown in Figure~\ref{fig:time}(b), PACT is consistently cheaper while also achieving higher $F_{\max}$ on 32 of the 35 shared designs. This efficiency comes from PACT's constrained backend harness. PACT uses structured actions, validation-gated acceptance, rollback and a within-run case record. These mechanisms avoid repeated unsafe or ineffective tool exploration. By contrast, the free-form Codex Agent repeatedly re-derives context and re-validates trial-and-error actions. This raises token consumption without matching PACT's timing gains. DATuner has no token cost (it is not LLM-based) but pays for search through substantially longer wall-clock runtime (Figure~\ref{fig:time}(a)).

\subsection{Ablation and Generalization Analysis}
\label{sec:ablation}

\begin{table}[t]
\centering
\caption{Geometric-mean $F_{\max}$ improvement (\%) on the 27
development and 8 held-out test designs. \emph{Scripted} replaces the
LLM with PACT's rule layer over the same actions, gates and budget;
\emph{phys\_opt} is a pure Vivado directive sweep with no agent.}
\label{tab:ablation}
\begin{tabular}{lrrr}
\toprule
Method & Dev (27) & Test (8) & All (35) \\
\midrule
PACT (full)            & $\mathbf{+27.70}$ & $+5.70$ & $\mathbf{+22.30}$ \\
\;\;Scripted (no LLM)  & $+16.78$ & $+5.54$ & $+14.11$ \\
\;\;phys\_opt sweep    & $+3.53$  & $+1.12$ & $+2.98$  \\
DATuner                & $+18.39$ & $+4.80$ & $+15.14$ \\
Codex Agent            & $+12.70$ & $+0.47$ & $+9.78$  \\
\bottomrule
\end{tabular}
\end{table}

We split the 35 designs into 27 development and 8 held-out test designs
that the agent optimizes cold, and add two ablations
(Table~\ref{tab:ablation}). A pure \texttt{phys\_opt\_design} directive
sweep, Vivado's own post-route optimizer, with no agent or
RapidWright, reaches only $+2.98\%$, far below PACT, so the commercial
backend alone cannot express the local edits PACT applies. Replacing
the LLM with PACT's deterministic rule layer (\emph{Scripted}) gives
$+14.11\%$: agentic action selection supplies the remaining gap, mostly
on development designs. On the held-out designs PACT ($+5.70\%$) and
Scripted ($+5.54\%$) are comparable and both exceed the uncapped
DATuner ($+4.80\%$), so the structured action set and validation
harness, rather than the LLM, carry generalization to unseen
checkpoints.

We further run a leave-one-out test to measure cross-design evidence transfer. The test uses five designs from three families: Verilog CORDIC-L/M, Double-Clock FFT-L/M, and the held-out NVDLA Large. For each target design, we compare three settings: no prior evidence, evidence from family members, and evidence from the target design itself. The final $F_{\max}$ values show little variation across the three settings; for example, CORDIC-L reaches (+26.7\%) in each setting.

\rev{These similar outcomes suggest that the transferable cases add
little decision value beyond the recipe planner. For these designs the
useful cross-design knowledge is compact: once the profile exposes the
dominant bottleneck, the curated recipe and the retrieved cases favor
the same action. The case record then mainly prunes repeated or invalid
trials within a run and keeps the trajectory auditable. Larger
cross-design gains likely require finer EDA-report semantics, for
example when several RapidWright ECOs fit one profile; such cases are
rare in this suite. Overall, PACT's gains come mainly from the recipe
planner, the structured action set and validation-gated execution.}

\section{Conclusion}
\label{sec:conclusion}

This paper presented PACT, a post-route agentic checkpoint-tuning
framework that operates on
implemented Vivado checkpoints \rev{through} structured
Vivado and RapidWright actions. Each candidate edit is a
validation-gated transition. PACT commits it only when it improves
timing while preserving routing legality, hold and pulse-width safety
and checkpoint usability, and rolls it back otherwise; functional
equivalence holds by construction. Across 35 UltraScale+ post-route checkpoints PACT improves
validation-clean $F_{\max}$ by a geometric mean of $+22.30\%$, ahead of
DATuner ($+15.14\%$) and the Codex Agent ($+9.78\%$), while running
$6.4\times$ faster than DATuner and at $24.5\times$ lower token cost
than the Codex Agent.

The results suggest that post-route timing closure is limited not only
by search time but also by action granularity and validation
discipline. \rev{PACT's gains come from local checkpoint edits and a
recipe planner beyond the reach of global search. Future work will
extend PACT to cross-design evidence transfer, multi-clock, cross-die
and power-aware optimization.}

\section*{Acknowledgment}
The work is supported by 
the AI for Science Program, Shanghai Municipal Commission of Economy and Informatization (2025-GZL-RGZN-BTBX02038), 
and Shanghai Anlogic Infotech Co. Ltd..

\newpage


\end{document}